\documentclass[aps,twocolumn,showpacs,superscriptaddress,amsmath,amssymb,amsfonts,floatfix,longbibliography]{revtex4-1}

\usepackage[T1]{fontenc} 
\usepackage{graphicx}
\usepackage{float}

\usepackage{dcolumn}
\usepackage{bm}
\usepackage{amssymb}
\usepackage{microtype}
\usepackage{xfrac}
\usepackage{gensymb}
\usepackage[most]{tcolorbox}
\usepackage{xcolor}
\usepackage{multirow}
\usepackage{enumitem}
\usepackage{natbib,hyperref}
\usepackage{graphicx}
\usepackage{booktabs}

\hypersetup{
	colorlinks=true, 
	citecolor=blue,  
}

\begin{document}

\title{Intrinsic defect intolerance in the ultra-pure metal PtSn$_4$}

\author{Samikshya Sahu}
    \affiliation{Stewart Blusson Quantum Matter Institute, University of British Columbia, Vancouver, BC V6T 1Z4, Canada}
    \affiliation{Department of Physics \& Astronomy, University of British Columbia, Vancouver, BC V6T 1Z1, Canada}

\author{Dong Chen}
    \affiliation{Stewart Blusson Quantum Matter Institute, University of British Columbia, Vancouver, BC V6T 1Z4, Canada}
    \affiliation{Department of Physics \& Astronomy, University of British Columbia, Vancouver, BC V6T 1Z1, Canada}

\author{Niclas Heinsdorf}
    \affiliation{Stewart Blusson Quantum Matter Institute, University of British Columbia, Vancouver, BC V6T 1Z4, Canada}
    \affiliation{Department of Physics \& Astronomy, University of British Columbia, Vancouver, BC V6T 1Z1, Canada}
    \affiliation{Max Planck Institute for Solid State Research, Heisenbergstrasse 1, 70569 Stuttgart, Germany} 
    
\author{Ashley N. Warner}
    \affiliation{Stewart Blusson Quantum Matter Institute, University of British Columbia, Vancouver, BC V6T 1Z4, Canada}
    \affiliation{Department of Physics \& Astronomy, University of British Columbia, Vancouver, BC V6T 1Z1, Canada}
    
\author{Markus Altthaler}
    \affiliation{Stewart Blusson Quantum Matter Institute, University of British Columbia, Vancouver, BC V6T 1Z4, Canada}
    \affiliation{Department of Physics \& Astronomy, University of British Columbia, Vancouver, BC V6T 1Z1, Canada}

\author{Ashutosh K. Singh}
    \affiliation{Stewart Blusson Quantum Matter Institute, University of British Columbia, Vancouver, BC V6T 1Z4, Canada}
    \affiliation{Department of Physics \& Astronomy, University of British Columbia, Vancouver, BC V6T 1Z1, Canada}
    
\author{Douglas A. Bonn}
    \affiliation{Stewart Blusson Quantum Matter Institute, University of British Columbia, Vancouver, BC V6T 1Z4, Canada}
    \affiliation{Department of Physics \& Astronomy, University of British Columbia, Vancouver, BC V6T 1Z1, Canada}

\author{Sarah A. Burke}
    \affiliation{Stewart Blusson Quantum Matter Institute, University of British Columbia, Vancouver, BC V6T 1Z4, Canada}
    \affiliation{Department of Physics \& Astronomy, University of British Columbia, Vancouver, BC V6T 1Z1, Canada}
    \affiliation{Department of Chemistry, University of British Columbia, Vancouver, British Columbia, Canada V6T 1Z1}
    
\author{Alannah M. Hallas}
    \email{alannah.hallas@ubc.ca}
    \affiliation{Stewart Blusson Quantum Matter Institute, University of British Columbia, Vancouver, BC V6T 1Z4, Canada}
    \affiliation{Department of Physics \& Astronomy, University of British Columbia, Vancouver, BC V6T 1Z1, Canada}
    \affiliation{Canadian Institute for Advanced Research (CIFAR), Toronto, ON, M5G1M1, Canada}

\date{\today}

\begin{abstract}

Ultra-pure materials are highly valued as model systems for the study of intrinsic physics. Frequently, however, the crystal growth of such pristine samples requires significant optimization. PtSn$_4$ is a rare example of a material that naturally forms with a very low concentration of crystalline defects. Here, we investigate the origin of its low defect levels using a combination of electrical resistivity measurements, computational modeling, and scanning tunneling microscopy imaging. While typical flux-grown crystals of PtSn$_4$ can have residual resistivity ratios (RRRs) that exceed 1000, we show that even at the most extreme formation speeds, the RRR cannot be suppressed below 100. This aversion to defect formation extends to both the Pt and Sn sublattices, which contribute with equal weight to the conduction properties. Direct local imaging with scanning tunneling microscopy further substantiates the rarity of point defects, while the prohibitive energetic cost of forming a defect is demonstrated through density functional theory calculations. Taken together, our results establish PtSn$_4$ as an intrinsically defect-intolerant material, making it an ideal platform to study other properties of interest, including extreme magnetoresistance and topology.

\end{abstract}

\maketitle

\section{\label{sec:Introduction}Introduction}

In the crystal grower's toolkit, the residual resistivity ratio (RRR~$ = \rho_{300\rm{K}}/\rho_0$) is a shorthand for assessing the quality of metallic samples. For normal metals, the resistivity in the 0~K limit, $\rho_0$, is intrinsically limited by the concentration of defects and other forms of disorder that disrupt the otherwise perfect periodic potential. As the temperature increases, the resistivity is dominated by thermal effects, where atomic vibrations of increasing amplitude result in progressively larger disruptions to the periodic potential. For a given material, the resistivity at room temperature  $\rho_{300\rm{K}}$ is therefore relatively sample independent whereas $\rho_0$ strongly depends on sample quality. Consequently, the most pristine and disorder-free samples are those with the highest RRR values. These samples are highly valued, particularly in the case of so-called quantum materials, whose electronic and magnetic states are known to be highly susceptible to extrinsic defect-driven effects~\cite{liang2015ultrahigh,ali2015correlation,campbell2021topologically,zhang2024crystal}.

The process of optimizing a crystal growth to maximize the RRR is often labor-intensive and iterative, typically involving multiple growth methods and taking place over years or even decades. A topical example comes from the exotic actinide superconductor, UTe$_2$. Crystals of UTe$_2$ grown by the tellurium self-flux method can have such low crystal quality (RRR$ < 2$) that in some instances no superconducting transition is detected ~\cite{aoki2019unconventional,ran2021comparison,sakai2022single}. Meanwhile, crystals grown by vapor transport (RRR~$ < 100$) are highly variable in their superconducting transition temperatures $T_c$~\cite{rosa2022single}, with some samples exhibiting multiple transitions ~\cite{hayes2021multicomponent} or spatial inhomogeneity ~\cite{thomas2021spatially}.
The most optimized crystals grown by a metathesis molten salts method (RRR~$ > 1000$) exhibit the highest observed $T_c$ values and a single well-defined transition under otherwise ambient conditions~\cite{sakai2022single,wu2024enhanced}. This saga has unfolded over the past six years, and at each stage, the improvement in crystal quality has changed the interpretation of this material's underlying physics. 
Similar stories have played out in a host of quantum materials, including cuprates \cite{liang2012growth,erb1996use}, and ruthenates~\cite{bobowski2019improved,perry2004systematic}, among many others~\cite{kim2015crystal,chen2020unconventional,krellner2012single}. Materials optimization problems are also common for semiconductors and insulators, but without the convenient shorthand of RRR to guide the way. In the case of semiconductors, carrier mobility can serve as a similar proxy to RRR, while insulators often require more cumbersome direct measurements.

\begin{figure*}[htbp]
  \centering
  \includegraphics[width=\linewidth]{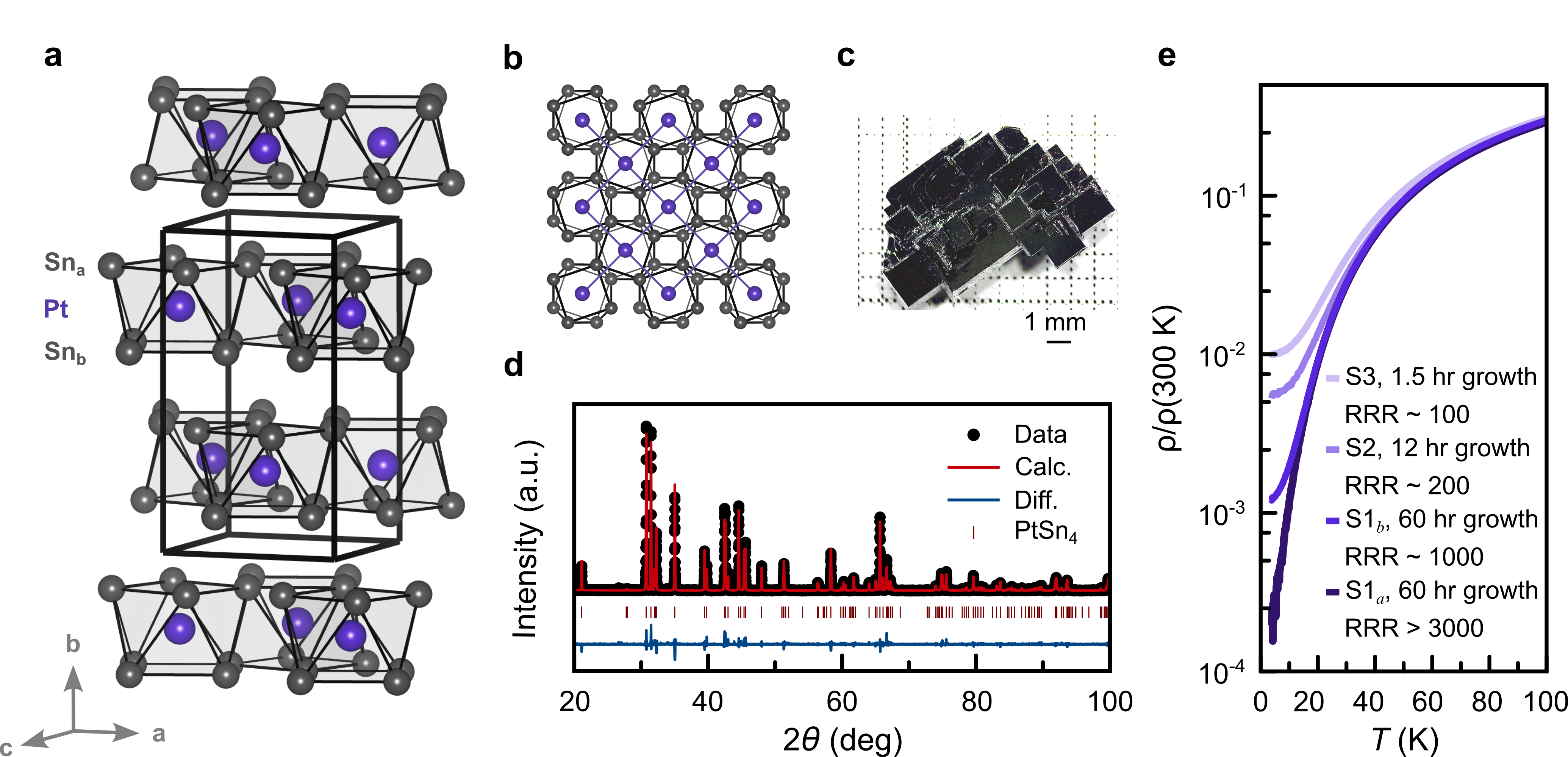}
  \caption{\textbf{Structural and electrical properties of PtSn$_4$.} (a,b) The quasi two-dimensional crystal structure of PtSn$_4$ is composed of staggered layers, where each layer is composed of a Pt square lattice sandwiched between two layers of Sn. Within each layer, square antiprismatic PtSn$_8$ units have an edge-sharing coordination as can be seen when viewing the structure along the $b$-axis. Sn$_\text{a}$ and Sn$_\text{b}$ label two crystallographically equivalent positions that when exposed as the top layer through cleaving yield distinct surfaces. (c) The metallic self-flux method yields cm-scale crystals of PtSn$_4$ with well-defined rectangular facets along the crystallographic $b$-axis. The as-grown crystal is found to cleave easily in this direction. (d) Rietveld refinement of powder x-ray diffraction data confirms the orthorhombic \emph{Ccca} (no. 68) structure and high crystalline quality. (e) These as-grown crystals of PtSn$_4$ exhibit remarkable residual resistivity ratios (RRRs) above 1000 for crystals grown with a standard cooling rate (60 hrs). Dramatically accelerated cooling (12 and 1.5 hrs) still yields crystals with RRR values over 100, indicative of very good crystal quality and very low levels of defects.}
  \label{fig:structure}
\end{figure*}

Rarely, some materials exhibit robust and remarkably high RRR values without significant optimization of the crystal growth process. One such material is the multi-band metal PtSn$_4$, which consistently exhibits RRR values approaching and even exceeding 1000~\cite{mun2012magnetic,luo2018origin,yan2020giant}. PtSn$_4$ is a multifaceted material, having attracted interest for an array of properties including its ``extremely large'' magnetoresistance (XMR)~\cite{mun2012magnetic,luo2018origin,yan2020giant}, its putative topological state~\cite{wu2016dirac,wang2018topological,yan2020giant,lin2024ultrafast}, its non-diffusive electrical transport phenomena~\cite{fu2020largely}, and even its catalytic activity~\cite{li2019dirac}. The relationship between these disparate physical phenomena, if any, remains unclear. However, it has recently been demonstrated that the XMR can be understood semi-classically without invoking any more exotic theories~\cite{diaz2024semi}.

In this work, we investigate the origin of the unique electrical transport behavior in PtSn$_4$. Through deliberate attempts to create higher defect concentrations in the crystal growth process through extreme cooling rates, we are unable to produce crystals of PtSn$_4$ with RRR values of \emph{less than} 100. Chemical substitution, on the other hand, is found to be very effective at generating defects with IrSn$_4$ and AuSn$_4$ showing factors of three and ten reductions in RRR, respectively, under optimal growth conditions. We directly quantify the concentration of defects in PtSn$_4$ with scanning tunneling microscopy using large-scale imaging surveys, finding an astonishingly low level of defects in the Pt layers of the as-grown crystals. These findings are corroborated by density functional theory (DFT) calculations that confirm the steep energetic penalty for forming a defect in this material. 
We conclude that PtSn$_4$ is a solid state analog to the Goldilocks fairytale, wherein its structure is ``just right'', resulting in natural defect-intolerance.

\section{Results and Discussion}

\subsection{PtSn$_4$ as an intrinsically defect-free metal}

PtSn$_4$ is a layered material that crystallizes in the centrosymmetric orthorhombic $Ccca$ space group (no. 68). Each layer in this structure is comprised of a perfectly two-dimensional Pt layer that is sandwiched between two perfectly two-dimensional Sn layers. These Sn-Pt-Sn slabs are layered in an $ABAB$ stacking sequence as shown in Fig.~\ref{fig:structure}(a). 
The Pt atoms in PtSn$_4$ form a slightly distorted square lattice and each Pt sits inside an 8-fold coordinate square anti-prism of Sn. The Sn-Pt-Sn sandwiches are formed by edge-sharing PtSn$_8$ units as seen from the out-of-plane direction shown in Fig.~\ref{fig:structure}(b). While this structure is orthorhombic, the distortion from tetragonal symmetry is very minor, with the in-plane lattice parameters differing by only 0.6\%.

Large single crystals of PtSn$_4$ are readily obtained following a standard self-flux method using an excess of Sn as the flux. The obtained crystals often grow to the maximum size of the crucible and are highly faceted and shiny with minimal visible imperfections, as shown for the centimeter-scale representative crystal in Fig.~\ref{fig:structure}(c). The large rectangular facets point along the stacking direction, which is the crystallographic $b$-axis, and the crystals readily cleave along this direction. Although PtSn$_4$ has been the subject of several comprehensive investigations in recent years~\cite{mun2012magnetic,wu2016dirac,wang2018topological,fu2020largely,yan2020giant,lin2024ultrafast}, its structural characterization with standard x-ray diffraction methods has proven challenging. PtSn$_4$ crystals are highly malleable, and when attempting to mechanically break or crush the material to obtain a specimen for either single crystal or powder diffraction, they are prone to deformation rather than cleanly breaking. As a result, the measured Bragg peaks are significantly broadened despite the high quality of the single crystal specimens~\cite{mun2012magnetic,fu2020largely}. 
Here, we address this issue by applying an annealing protocol to our powder specimen following mechanical grinding, yielding a resolution-limited x-ray powder diffraction pattern with minimal preferred orientation as presented in Fig.~\ref{fig:structure}(d). 
Rietveld refinement gives excellent agreement with the orthorhombic $Ccca$ (no. 68) space group. A comparison of powder x-ray diffraction data taken pre- and post-annealing of the samples, along with the tabulated fitting parameters from the Rietveld refinement, is shown in Supplementary Fig.~S1 and Supplementary Table~S1.

The longitudinal in-plane electrical resistivity $\rho$(T), measured as a function of temperature from 2 K to 300 K, is the first screening probe we use to investigate the quality of our PtSn$_4$ crystals. This resistivity is measured using a four-terminal lead geometry on the \textit{ac} plane, where the geometry of the probe compensates for the contact resistance of the leads. From this measurement, we can extract the residual resistivity ($\rho$$_0$) and the residual resistivity ratio (RRR). 
As our crystals were grown from excess Sn flux, we define $\rho_0$ at 3.8 K to avoid signal contamination from the superconducting transition of Sn ($T_c = 3.6$~K), which could be observed in some of our samples due to minor Sn inclusions between the layers during the growth. Likewise, we define RRR as $\rho$(300 K)/$\rho$(3.8 K). 
Our typical flux-grown crystals of PtSn$_4$ have residual resistivities on the order of $\rho_0 =$~10--100 n$\Omega$-cm and RRRs close to 1000. 
The $\rho$(T) data for an assortment of typical PtSn$_4$ samples measured with varying sample preparation protocols are shown in Supplementary Fig.~S2, giving RRR values that vary from 785 to above 3000. These remarkable transport properties indicate excellent conductivity with very low defect concentrations and are obtained without any crystal growth optimization, in striking contrast to the usual process of iteratively improving the crystalline quality.

Our observations of low defect concentrations in PtSn$_4$ are not an anomaly: previous studies on its electrical transport properties have found RRR values varying from 700 to 1000~\cite{mun2012magnetic,luo2018origin,li2019dirac,fu2020largely,yan2020giant}.
To better understand the origin of these low scattering rates, we attempted to deliberately introduce higher concentration of defects in PtSn$_4$ by increasing the cooling rate during flux growth, where 
faster cooling is expected to thermodynamically trap higher number of defects as the crystal forms. Fig.~\ref{fig:structure}(e) shows the normalized resistivity with respect to room temperature resistivity ($\rho/\rho (300K)$) for four PtSn$_4$ growths with varying cooling times. Sample 1 (S1$_a$ and S1$_b$) is representative of our above-mentioned typical PtSn$_4$ samples and is cooled from the highest dwelling temperature to the spinning temperature with a 60 hr (4 \degree C/hr) cooling time; this is followed by sample 2 (S2), with a 12 hr (20 \degree C/hr) cooling time, which is then followed by sample 3 (S3) with a 1.5 hr (165 \degree C/hr) cooling time. To put these cooling rates into perspective, flux growth is typically performed with cooling rates between 1 and 10 \degree C/hr~\cite{kanatzidis2005metal,canfield2010solution} and accordingly we define S1 as our typical crystal of PtSn$_4$. By comparison, the 20\degree C/hr cooling rate of S2 is fast and the 165\degree C/hr cooling rate of S3 is exceptionally fast. In fact, the cooling rate for S3 was the maximum achievable with our furnace. The $\rho$$_0$ and RRR values for these four samples are given in Table \ref{Table 1}, where the highest quality S1$_a$ has a RRR above 3000 while the fast cooled sample S3 has a RRR of 100. While the decreasing RRR indicates that we have successfully formed a larger concentration of defects, it is important to emphasize that a RRR of 100 would be considered to indicate excellent crystalline quality in most materials. Even with air quenching, we were unable to prepare a sample of PtSn$_4$ with a RRR of less than 100. We therefore conclude that  PtSn$_4$ is naturally intolerant of forming defects as it crystallizes.

\begin{table}[htbp]
  \caption{Measured values for RRR and $\rho_0$ at 3.8 K for four PtSn$_4$ crystals that were grown at different cooling rates.}
  \begin{ruledtabular}
  \renewcommand{\arraystretch}{1.2}
    \begin{tabular}{lccc}
    Sample & Cooling rate~(\degree C/hr) & RRR   & $\rho_0$~(n$\ohm$-cm) \\
    \colrule
    S1$_a$ & 4  & $>$\,3000   & 16 \\
    S1$_b$ & 4  & 835   & 91 \\
    S2  & 20 & 178   & 204 \\
    S3  & 165 & 100   & 504 \\
    \end{tabular}
    \end{ruledtabular}
  \label{Table 1}
\end{table}

\begin{figure*}[htbp]
  \centering
  \includegraphics[width=\linewidth]{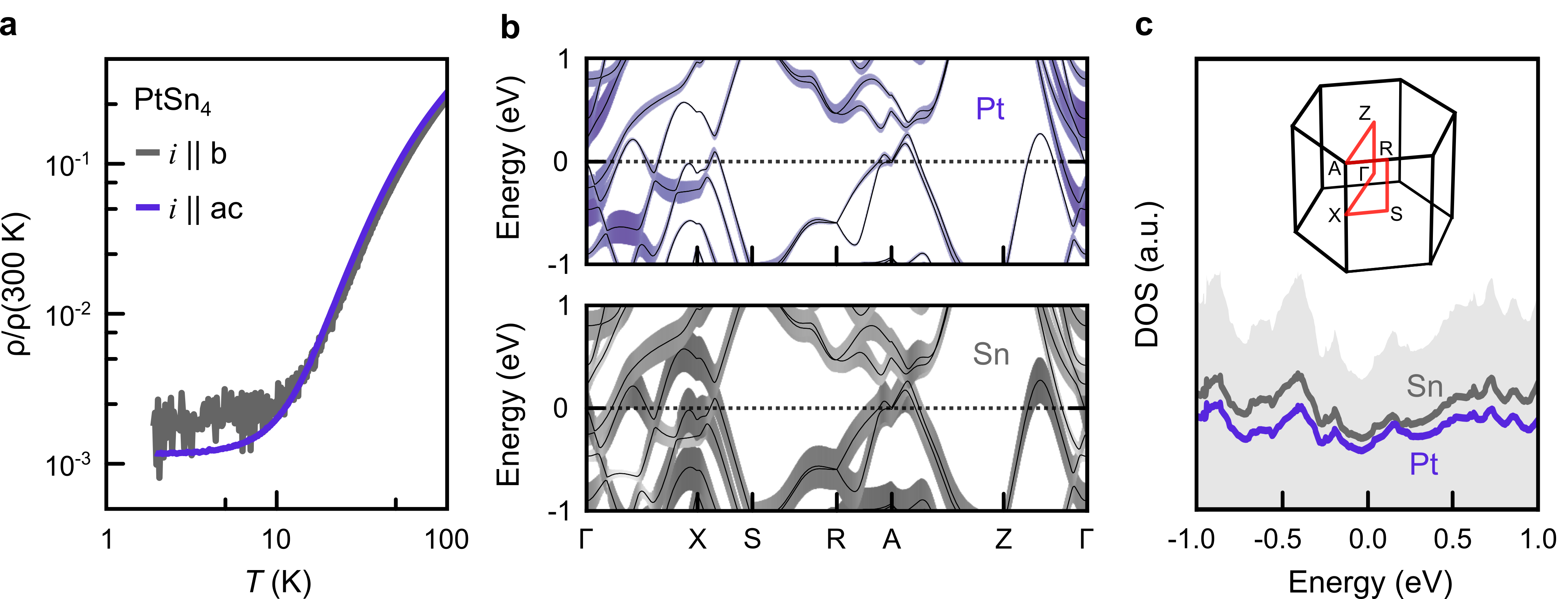}
  \caption{\textbf{The role of dimensionality in PtSn$_4$.
  }(a) Normalized resistivity as a function of temperature for in-plane (\textit{i} $\parallel$ ac) and out-of-plane (\textit{i} $\parallel$ b) directions in as-grown PtSn$_4$ crystals plotted on a log-log scale. The absolute value of resistivity for the two directions is similar in magnitude, indicative of relatively isotropic electrical conductivity, typical for a three-dimensional material. For (\textit{i} $\parallel$ b) the data below 10 K are at the signal detection threshold.   (b) Projected band structures onto the Pt (purple) and Sn atoms (gray). (c) DOS with contributions from the Pt (purple) and Sn (gray). At the Fermi level, the total Sn weights are only slightly higher than the Pt weights. This demonstrates that both Pt and Sn layers contribute to the overall transport in PtSn$_4$. The insert shows the first BZ of the primitive unit cell of PtSn$_4$ and the chosen high-symmetry path (red). }
  \label{fig:2}
\end{figure*}

\subsection{Role of dimensionality in the electronic properties of PtSn$_4$}

We next consider the role of dimensionality, and specifically, the role of individual Pt and Sn layers, in determining the transport properties of PtSn$_4$. This line of inquiry is motivated by comparison with other highly conductive and structurally anisotropic metals, such as PdCoO$_2$~\cite{shannon1971chemistry,tanaka1996growth,ong2010unusual,hicks2012quantum,bachmann2022directional} and Sr$_2$RuO$_4$~\cite{lichtenberg1992sr2ruo4,maeno1997two,tyler1998high}. The transport behavior in these materials is highly anisotropic, with the in-plane and out-of-plane resistivities varying by multiple orders of magnitude, as a direct consequence of their relatively simple two-dimensional Fermi surfaces~\cite{bergemann2003quasi,eyert2008metallic,hicks2012quantum}. Furthermore, in the case of PdCoO$_2$, the high in-plane conductivity is directly attributable to the segregation of defects onto the insulating Co-O layers, which leave the conductive Pd layers pristine~\cite{zhang2024crystal}. It is important to emphasize that although PtSn$_4$ is also structurally anisotropic, it is known to possess a complex, multiband, and highly three-dimensional Fermi surface~\cite{mun2012magnetic,yara2018small,diaz2024semi}.

To investigate whether the Pt or Sn sublattices in PtSn$_4$ may be uniquely responsible for the very high conductivity, we first consider the anisotropy of the electrical resistivity. Through inspection of the crystal structure of PtSn$_4$, we can see that conduction along the out-of-plane direction must naturally involve both metallic sublattices, whereas in-plane transport could be dominated by a single sublattice. In-plane (\textit{i} $\parallel$ ac) and out-of-plane (\textit{i} $\parallel$ b) transport measurements therefore provide qualitative evidence of the relative defect concentrations of the two sublattices. The samples used for this characterization are representative of the typical S1 samples mentioned in the previous section. The in-plane measurement was performed via a standard four-terminal method, while the out-of-plane measurement employed a modified four-terminal geometry as described in the methods~\cite{bud1998anisotropic}. The resulting data, normalized by the room temperature value ($\rho/\rho (300K)$), is shown on a log-log scale in Fig.~\ref{fig:2}(a). A small deviation is noted at the lowest measured temperatures as the resistivity saturates at a slightly larger value for the \textit{i} $\parallel$ b direction. However, as evidenced by the overall noise level, this measurement surpasses the signal detection limit for the lock-in amplifier in this orientation below 10 K. Recent work on microstructured samples has established that PtSn$_4$ has minimal anisotropy in its resistivity at room temperature~\cite{zhang2024crystal}, so the data shown here indicates there also very little anisotropy in either the temperature dependence or in the residual resistivity at low temperatures. We, therefore, conclude that PtSn$_4$ is an isotropic 3D conductor with both Pt and Sn layers taking part in conduction and a low level of defects on both sublattices.

In support of the above claim, Fig. \ref{fig:2}(b) shows the band structure of PtSn$_4$ calculated with density functional theory (DFT) along a path through the base-centered orthorhombic Brillouin zone (BZ), which is shown in the inset to Fig. \ref{fig:2}(c). The top and bottom panels show the projections onto the Pt and Sn atoms, respectively, where the thickness and intensity of the color represent the weighted atomic contribution. PtSn$_4$ is clearly metallic with a complicated, multi-sheeted Fermi surface that consists of four bands crossing the Fermi level forming two hole and two electron pockets. Although PtSn$_4$ has a layered crystal structure, the strongly dispersive bands along the $Z$-$\Gamma$ path confirm that this material is not electronically two-dimensional which is consistent with the lack of anisotropy in the electrical transport. The bands in the presented $k$ path are predominantly of Sn character. However, the total density of states (DOS) in the window of energies close to the Fermi level is only slightly larger for Sn than that of the Pt atoms, as shown in \ref{fig:2}(c).

We can therefore conclude on both experimental and computational grounds that neither of the sublattices in PtSn$_4$ dominate its electrical transport properties, but rather they both contribute with approximately equal weights. In making sense of this result we can return to the crystal structure shown in Fig.~\ref{fig:structure}(a,b). The shortest interatomic distance in this orthorhombic structure are the Pt-Sn bonds in the square antiprism units that make up the Sn-Pt-Sn sandwich, which are 2.78 and 2.79~\AA\ at room temperature. The next shortest interatomic distance is between adjacent Sn layers, where the Sn-Sn bonds are just slightly larger at 3.00~\AA, which is actually shorter than the Sn-Sn distances in elemental Sn at room temperature \cite{wolcyrz1981x}. It is therefore clear that there is strong hybridization both within the plane and along the stacking direction. Returning to our questions of the natural defect intolerance observed in PtSn$_4$, we can now establish that this property must extend to both the Pt and Sn sublattice.

\subsection{Electrical transport and defect formation trends in the $M$Sn$_4$ family}

\begin{figure}[tbp]
  \centering
\includegraphics[width=\linewidth]{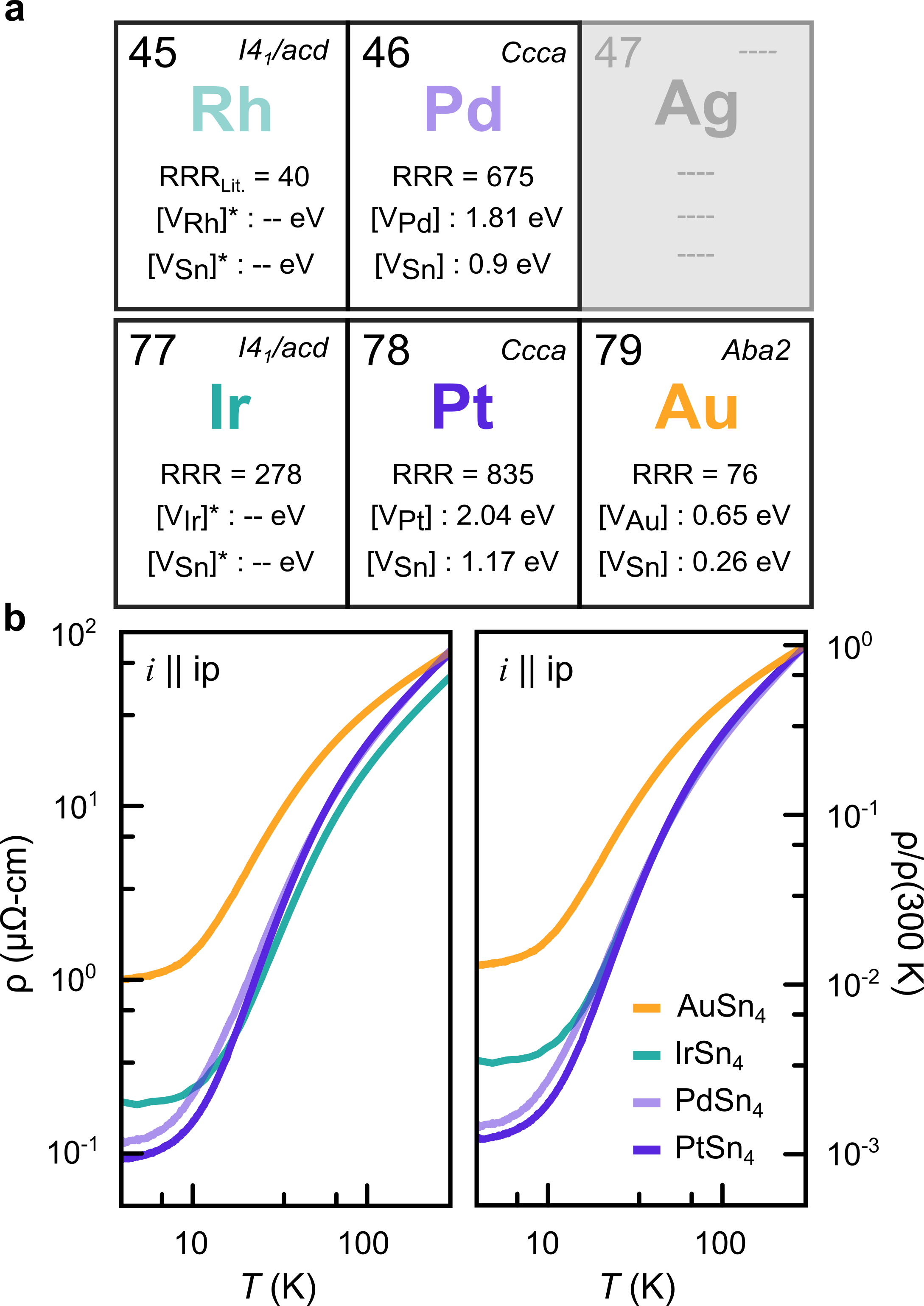}
  \caption{\textbf{Comparison of the structure-property relationships within comparable members in the \textit{M}Sn$_4$ family.} (a) Illustrates a periodic table representation of the members of the \textit{M}Sn$_4$ (\textit{M}= Rh, Pd, Ir, Pt, or Au) family showing the respective space group of formation, along with experimentally determined residual resistivity ratio (RRR), and DFT calculated vacancy defect formation energy for \textit{M} and Sn for the relevant crystal structure. The RRR value for RhSn$_4$ is taken from literature \cite{xing2016large}. An Ag analog, indicated by the grayed-out rectangle, does not exist. (b) Temperature dependence of the in-plane (ip) resistivity (left panel) and resistivity normalized to room temperature (right panel) for AuSn$_4$ (yellow), IrSn$_4$ (teal), PdSn$_4$ (light purple), and PtSn$_4$ (purple). While the room temperature resistivity values are comparable across the four materials, the residual resistivity at base temperature differs by more than an order of magnitude.}
  \label{fig:3}
\end{figure}

Next, we seek to understand the relationship between the defect intolerance of PtSn$_4$ and its chemical composition. The family of materials to which PtSn$_4$ belongs is structurally complex. Current reports identify that, with the exception of PdSn$_4$, all other chemical substitutions, at both the Pt site and the Sn site yield \emph{nearly} isostructural materials with differing space group symmetries. Here we specifically consider the case of chemical substitution of Pt by its periodic table neighbours, Pd, Ir, and Au, which sit directly above, and to the left, and right of Pt in the periodic table, respectively, as shown in Fig.~\ref{fig:3}(a). PdSn$_4$ alone is isoelectronic to PtSn$_4$ and crystallizes in the \textit{Ccca} space group~\cite{kubiak1984refinement, nylen2004structural, jo2017extremely}. RhSn$_4$ and IrSn$_4$ are reported to crystallize in the \textit{I4$_1$/acd} space group (no. 142)~\cite{nordmark2002polymorphism, xing2016large} while AuSn$_4$ crystallizes in the \textit{Aba2} space group (no. 41)~\cite{kubiak1984refinement, yadav2025low}, which are tetragonal and orthorhombic, respectively. Interestingly, the basic building block in all these structures is the edge sharing square antiprismatic unit $M$Sn$_8$ ($M=$~Rh, Pd, Ir, Pt, and Au), as shown in Fig.~\ref{fig:structure}(b). In all these cases, the $M$ site forms a slightly distorted (undistorted in the case of $M=$~Rh or Ir) square lattice that has an $ABAB$ ($ABCD$ in case of $M=$~Rh and Ir) type stacking along the out-of-plane direction. The structural differences between these five materials originate from subtle stacking variations at the Sn site~\cite{herrera2023band}.

When comparing the formation of PtSn$_4$ to its Au and Ir analogs, there are two interesting observations that are suggestive of a Goldilocks type effect. 
The first of these observation is for IrSn$_4$, where the nearly isostructural phase described here (so called $\beta$-IrSn$_4$) is metastable and can only be stabilized at high temperatures~\cite{nordmark2002polymorphism} or high pressures~\cite{larchev1984new}. The low temperature $\alpha$-IrSn$_4$ phase~\cite{lang1996transition}, which is found to be energetically preferred in DFT calculations~\cite{nordmark2002polymorphism}, is trigonal with no obvious resemblance to the $\beta$ phase. The second observation is for AuSn$_4$, which is observed to form with a high frequency of stacking sequence faults~\cite{kubiak1981influence,kubiak1985x}. These stacking faults have even led to conflicting space group assignments: recent ARPES measurements are found to be more consistent with the $Ccca$ space group (no. 68)~\cite{herrera2023band} while electron diffraction favors $Aba2$ (no. 41)~\cite{zhu2023intrinsic}. Stacking faults due to polymorphism are also prevalent for the Pb-substituted variant, PtPb$_4$, with a reported $P4/nbm$ space group (no. 125)~\cite{rosler1951kristallstruktur,lee2021evidence,wu2022nonsymmorphic}. PtSn$_4$, however, is thermodynamically stable and free from a high concentration of stacking faults, suggestive of the higher stability of its structure relative to its substituted Ir and Au analogs.

We next compare results from electrical transport for the chemically substituted $M$Sn$_4$ family, wherein the Pt site is fully substituted by $M =$~Pd, Ir, or Au. Single crystals of PdSn$_4$, IrSn$_4$, and AuSn$_4$ were grown by a self flux method using comparable or higher purity starting reagents as in the case of our PtSn$_4$ growths. While each of these materials has a unique formation temperature and stability range within its binary phase diagram, we have attempted to keep the total cooling time and cooling rate comparable, as described in the Methods. Representative in-plane resistivities and the same data normalized to room temperature ($\rho/\rho (300K)$) are presented in Fig. \ref{fig:3}(b).  Previous studies have established superconducting states in IrSn$_4$ and AuSn$_4$, with critical temperatures of $T_c \sim 0.9$~K and $T_c \sim 2.4$~K, respectively \cite{tran2013observation, ahmad2024linear, shen2020two, zhu2023intrinsic, herrera2023band}. As our goal here is to use the normal state residual resistance as a probe of the intrinsic defect concentration, we truncate all our data sets at 3.8~K above both materials' superconducting transitions and above that of elemental Sn. All four samples are excellent conductors with very similar resistivities at room temperature, approximately 50 $\mu \Omega$-cm, while at low temperature there is an order of magnitude separation in their residual resistivities. The resistivities for these samples also show minimal anisotropy, as shown in Supplementary Fig.~S3. The RRR values for our representative IrSn$_4$ and AuSn$_4$ data sets are 278 and 76, respectively, which agree well with the previously reported literature values \cite{ahmad2024linear, zhu2023intrinsic}. Previous studies of chemical substitution on the Sn site with Pb yielding PtPb$_4$ also give RRR values of approximately 100 \cite{xu2021superconductivity,wang2021normal}. 

PdSn$_4$ alone stands out as comparable to PtSn$_4$ with its high RRR of 675. In addition to being isostructural and isoelectronic, Pd and Pt are significantly closer in atomic radii than Pt and Au or Ir. For example, using the metallic radii as a proxy, Pd and Pt differ by only 1\% while Pt and Au differ by 4\%. Consequently, the orthorhombic lattice parameters of PdSn$_4$ are in very close agreement with those of PtSn$_4$. Some literature reports have also found high RRRs for PdSn$_4$, including RRR = 331 and 625~\cite{jo2017extremely} and RRR = 408~\cite{zhang2019structural}. However, there are other reports of PdSn$_4$ crystals with significantly lower qualities, including RRR = 27~\cite{karn2023weak} and RRR = 81 and 87~\cite{xu2017enhanced}, which is lower than we were able to achieve in PtSn$_4$ through cooling at our maximum possible rate. We therefore conclude that although PdSn$_4$ shares many features with PtSn$_4$ due to their inherent structural and electronic similarities, PtSn$_4$ remains the superior realization of intrinsically low defect concentrations.

To gain insight into the naturally high RRR of PtSn$_4$, we employ DFT to study the vacancy formation in this material as a proxy for all point defects. While there are a multitude of point defects that can exist in a real material, including substitutions and interstitials, the cost of forming a vacancy is an excellent measure of the overall bond strength. Vacancies also come in a singular form, enabling straightforward comparison across different compositions and across both the $M$ and Sn sublattices. To contrast, in the case of substitutions, the ionic radius, valence, and chemical character of the specific substituent will significantly impact the energetic cost, making it difficult to standardize comparisons across the material family.
We calculate the formation energy of a vacancy $X$ as~\cite{formation_energy1, formationenergy2}
\begin{align}
    E^f[X] = E_{\text{tot}}[X] - E_{\text{tot}}[\text{bulk}] + \mu_X, 
\end{align}
with $E_{\text{tot}}[X]$ and $E_{\text{tot}}[\text{bulk}]$ being the total energies of supercells with and without a vacancy $X$, respectively. $\mu_X$ is the chemical potential that is derived from the ground state energy of the elementary phase of the vacancy species. The formation energy itself is mostly temperature-independent and can be accurately determined using first principle calculations at zero temperature. The temperature dependence of the defect concentration stems dominantly from the configurational entropy, which compensates the energy cost of creating a defect. The formation energies for vacancies at the noble metal ($M=$~Pd, Pt, Au) site and the Sn site, without posterior relaxation, along with the measured RRRs for the different compounds, are shown as a periodic table representation in Fig. \ref{fig:3}(a). These calculations cannot be reliably performed in the case of IrSn$_4$ due to its metastability~\cite{lang1996transition,nordmark2002polymorphism}. 

As PdSn$_4$, PtSn$_4$, and AuSn$_4$ are structurally similar and are grown in comparable temperature ranges, we expect the vacancy formation energy to be an indicator of the overall defect concentration and use it to qualitatively compare their relative abundances. A higher formation energy results in fewer vacancies and, hence, a higher RRR is expected.  
Notably, the energy cost of forming a Pt defect (2.04 eV) is roughly twice the cost of creating a Sn defect (1.17 eV) in PtSn$_4$, whereas the cost of forming an Au (0.65 eV) or Sn (0.26 eV) defect in AuSn$_4$ is significantly smaller. These findings suggest that the Sn layers in PtSn$_4$ will carry more defects than the Pt layers and that AuSn$_4$ is altogether more amenable to forming defects on both sublattices, consistent with the experimental findings from electrical transport measurements. As expected, the values for PdSn$_4$ are comparable but smaller than those for PtSn$_4$. In particular, considering defects on the Sn site, which are expected to be more prevalent, we find that Sn vacancy formation energy is 20\% lower in PdSn$_4$ as compared to PtSn$_4$, which is substantial when considering that the probability of forming a defect scales with a Boltzmann-like activation. In the case of IrSn$_4$, while the metastability inhibits our ability to study the vacancy formation cost, we can nonetheless conclude that the higher crystal growth temperature (approximately twice as high as the temperature range in which the other analogs form) would naturally contribute to the formation of higher defect concentrations.

Another way of probing the propensity for a material to form a defect is to consider the relaxation of a supercell with an introduced vacancy. The extent to which the formation energy is reduced is indicative of how much a given crystal is able to accommodate the vacancy by structural relaxation. For a material with a Goldilocks structure (\emph{i.e.} the structural parameters fall in an optimal range given all constraints) we expect that percentage change due to relaxation to be small because  vacancies would likely be more abundant if the material could easily compensate their energy cost by relaxation. Consistent with this scenario, we find that relaxation reduces the formation energy of a cation in AuSn$_4$ by about $45$\% percent, while the energetic cost of a Pt vacancy in PtSn$_4$ is only brought down by $9$\% percent.

Finally, we briefly comment on our attempts to partially substitute Pt with either Pd, Ir, or Au. For all tested compositions, we find that partial substitutions of any concentration on the Pt site result in an actual substitution level that is substantially smaller than the nominal stoichiometry. For example, flux growth with a 50:50 ratio of Au to Pt results in a final stoichiometry of Pt$_{0.93}$Au$_{0.07}$Sn$_4$ as determined from energy dispersive x-ray spectroscopy (EDX) measurements. Similar results are obtained for substitutions of Pt with Ir, or even Pd (see Supplementary Tables~S2 and S3) which shows reduced uptake despite being isoelectronic and having a similar radius. We computed the energy cost of substituting a single Pt atom with Pd to be 0.3 eV, an order of magnitude smaller than creating a Pt vacancy but still strongly disfavored at the synthesis temperature., These results highlight the stability of the Pt sublattice toward substitution.

\begin{figure*}[htbp]
  \includegraphics[width=\linewidth]{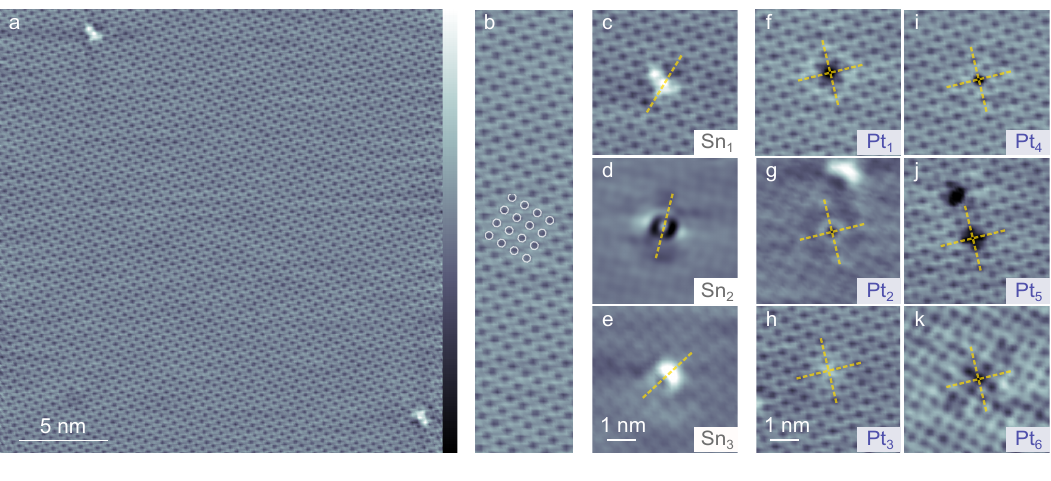}
  \caption{\textbf{Categorization of point defects in PtSn$_4$ via low-temperature scanning tunneling microscopy (STM).} (a) STM topography on pristine PtSn$_4$ surface ($\Delta$z = 30~pm) and (b) the Pt-lattice corrugation overlay shown as white outlined blue circles. Gallery of nine defect types observed in PtSn$_4$, including (c-e) three types of Sn-site defects and (f-k) six types of Pt-site defects. The black line on the defects indicates the corresponding mirror planes. Images are acquired at different sample bias and set point currents: (c,f,h,i,j) used (-400~meV, 500~pA), (d) used (400~meV, 600~pA), (e) used (-400~meV, 600~pA), (g) used (-500~meV, 500~pA), and (k) used (-400~meV, 500~pA). All measurements were performed at a temperature of 4.4~K.}
  \label{fig:STMtopos}
\end{figure*}

\subsection{Statistical analysis of defects in PtSn$_4$}

In order to gain further experimental insight into the defect chemistry at play in PtSn$_4$, we performed a microscopic defect analysis using scanning tunneling microscopy (STM). As a surface probe, STM enables a quantitative count of the surface defect density. Scanning on PtSn$_4$ consistently revealed large flat terraces with few step edges. Through a detailed termination analysis conducted in a step-edge region as shown in Supplementary Fig.~S4, we confirmed that Sn$_a$ is the dominant scanning terrace, this indicates that Sn$_a$ and Sn$_b$ is the most common cleaving plane. Our finding is aligned with a previous study that revealed the bonding nature in PtSn$_4$, with the  Sn-Sn bonds being weak covalent bonds and Pt-Sn bonds being metallic \cite{li2019dirac}. Other terminations corresponding to the Pt and Sn$_b$ layer were rarely observed, accounting for less than 1\% of the investigated area. Therefore, our defect statistics analysis is based on scans of the Sn$_a$ termination. 

Scanning on Sn$_a$ termination yields a weak corrugation consistent with the Sn layer structure and a square array of depressions corresponding to the Pt sites below, corresponding well with the Pt-Pt distance determined by diffraction. Pristine cleaved surfaces are remarkably defect-free over total scanning areas of approximately 10$^5$ atoms. Sparse defects were observed in STM topographies of both the slow-cooled (S1, RRR~$> 1000$) and fast-cooled (S2, RRR~$~200$) samples. In total, nine kinds of defects are observed. Defects were then assigned to Sn or Pt sites via location and local symmetry: approximately 2-fold for Sn and approximately 4-fold for Pt. Three types of Sn-site defects were observed consistently across both samples, while four types of Pt-site defects were observed in the high RRR samples, and five types were observed in the low RRR samples (see Fig.~\ref{fig:STMtopos}(c-j) and Table~\ref{tab:table1}). Defect densities for each type were acquired by counting the number of defects observed in large-area image surveys. Details of the methods for obtaining reliable defect densities and uncertainties are given in the Supplementary Information. It is worth noting that clustering of defects was not observed.

\begin{table*}
\begin{ruledtabular}
\renewcommand{\arraystretch}{1.5}  
\caption{Defect statistics of two samples of PtSn$_4$ grown at two different cooling rates: slow-cooled sample S1 (RRR~$\sim1000$) and fast-cooled sample S2 (RRR~$\sim200$).} \label{tab:table1}
\begin{tabular}{ccccc}
 Defect & Symmetry & Category & $\rho A_{\text{S1}}$ (per unit cell) & $\rho A_{\text{S2}}$ (per unit cell) \\ 
 \hline
 Sn$_1$ (Croissant)\footnote{This kind of defect is non-intrinsic} & 2-fold & Sn site & $1.00(0.02) \times 10^{-3}$ & $3.50(0.03) \times 10^{-3}$ \\
 Sn$_2$ (Spearhead)$^{\text{a}}$ & 2-fold & Sn site & $1.46(0.02) \times 10^{-3}$ & $0.216(0.007) \times 10^{-3}$ \\
 Sn$_3$ (Crescent)$^{\text{a}}$ & 2-fold & Sn site & $1.545(0.006) \times 10^{-4}$ & $0.033(0.008) \times 10^{-4}$ \\
 \hline
 Pt$_1$ & 4-fold & Pt site & $2.9(0.6) \times 10^{-5}$ & $5.6(0.5) \times 10^{-5}$ \\
 Pt$_2$ & 4-fold & Pt site & $3.8(0.6) \times 10^{-5}$ & $5.4(0.5) \times 10^{-5}$ \\
 Pt$_3$ & 4-fold & Pt site & $1.7(0.4) \times 10^{-5}$ & $2.1(0.4) \times 10^{-5}$ \\
 Pt$_4$ & 4-fold & Pt site & $2.8(0.5) \times 10^{-5}$ & Not Observed \\
 Pt$_5$ & 4-fold & Pt site & Not Observed & $2.0(0.2) \times 10^{-5}$ \\
 Pt$_6$ & 4-fold & Pt site & Not Observed & $0.5(0.1) \times 10^{-5}$ \\
 \hline
 Sn$_{\text{total}}$ &  &  & $2.61(0.03) \times 10^{-3}$ & $3.72(0.03) \times 10^{-3}$ \\
 Pt$_1$+Pt$_2$+Pt$_3$ &  &  & $0.84(0.09) \times 10^{-4}$ & $1.31(0.07) \times 10^{-4}$ \\
 Pt$_{\text{total}}$ &  &  & $1.1(0.09) \times 10^{-4}$ & $1.56(0.08) \times 10^{-4}$ \\
\end{tabular}
\end{ruledtabular}
\end{table*}

Defect densities for each type are reported in Table~\ref{tab:table1}. Pt site defects are significantly rarer than Sn site defects, as discussed below. Of the 6 types of Pt-site defects, the first three were observed in both samples. All Pt site defects are in the range of 10--50 ppm, and of the three types present in both high and lower RRR samples, all increase in the fast-cooled lower RRR sample. The overall density of Pt site defects increases for the lower RRR sample, consistent with the increased low-temperature resistivity expected to originate from point defect scatterers.

Defects residing on the Sn site are more prevalent in both samples, having densities that are between one and two orders of magnitude larger than the Pt site defects. However, the analysis is confounded by sensitivity to surface effects. Although significant increases in defects over time due to the adsorption of contaminants were not observed, other extrinsic variability appears to significantly influence the total Sn site defect density. In particular, the Sn$_1$ (croissant) defect is readily induced by the tip through voltage pulses or prolonged application of a bias with amplitude higher than 700meV, and these defects have been observed to move and even annihilate, as shown in Supplementary Fig.~S5, leading us to suspect these defects correspond to a Frenkel pair (a Sn adatom-Sn vacancy pair). The Sn$_2$ and Sn$_3$ defects anomalously decrease in the fast-cooled (lower RRR) sample, contrary to expectations from the transport measurements. We suspect these correspond to vacancies and/or adatoms that arise in an uncontrolled manner during cleaving, as the Sn surface is easily disturbed. With these extrinsic factors contributing to the total observed defect density, we can only place an upper limit on the number of scattering sites. However, we note that as no new defect types appear, and all are sensitive to disorder reorganization of atoms, we do not observe any Sn-site defects that would clearly correspond to chemical substitutions. In other words, while some level of intrinsic defects are surely present on the Sn site, they are dominated by extrinsic ones. 

The Pt site defects, however, provide a more reliable indicator of the very low intrinsic defect concentration. We can therefore examine the differences between samples S1 (high RRR) and S2 (lower RRR). Pt$_1$, as the dominant intrinsic defect, doubled in the lower RRR sample, and the Pt$_2$ defect also has a sizable increase in the low RRR sample, while Pt$_3$ does not show a notable change in the density. Taken in aggregate the increase in defects corresponds to a $1.5\times$ larger concentration of Pt site defects in S2 (lower RRR) as compared to S1 (high RRR). The increase in intrinsic defect densities is aligned with the decrease in the low-temperature resistivity across the two samples. However, naively, we would expect a linear correlation between defect concentration and residual resistivity due to electron-defect scattering. The observed increase of $1.5\times$ in the Pt site defects described here fails to account for the $5\times$ decrease in the RRR value for S2 as compared to S1. We therefore conclude that, although the intrinsic Sn site defect density is dramatically obscured by the surface effects described above, they must still account for the majority of increased scattering centers. This reasoning is consistent with both the higher density of Sn atoms relative to Pt and the significantly reduced cost to form a Sn site vacancy as compared to a Pt site vacancy as determined by DFT. Nonetheless, the remarkably small number of Pt site defects, where the occurrence of intrinsic defect is as rare as 1 in 10,000 unit cells, is good evidence that crystalline perfection is an intrinsic quality of PtSn$_4$.

\subsection{Connection with XMR and topology}

Having established PtSn$_4$ as a naturally defect-intolerant metal, we return to two of its other notable characteristics: its extreme magnetoresistance (XMR) and its reported topological character. In particular, PtSn$_4$ was the first material reported to host so-called Dirac node arcs on its surface, which are planar intersections of finite length~\cite{wu2016dirac}. Since its discovery only two other materials with similar boundary modes have been reported~\cite{nodearc1, nodearc2}, where the authors suggest the interplay of different bulk nodal loops might potentially play a role~\cite{nodearc2}. 
To our knowledge, these modes are not predicted by any bulk topological classification or even toy model. In the presence of SOC, there are no extended degeneracies in PtSn$_4$'s bulk band structure~\cite{orthorhombic}. Even expanding our scope to include higher symmetry structures, which can manifest topological signatures when the structural distortion is small relative to the topological gap~\cite{mollquasi, irf4, stableedgemodes}, we see that none of the minimal tetragonal supergroups for $Ccca$~\cite{ivantchev2000subgroupgraph, ivantchev2002supergroups} have the suitable nodal line structure hypothesized to manifest the observed Dirac node arcs~\cite{nodearc2}.  

Instead, we want to point out the striking resemblance of the Dirac node arcs to hourglass surface state that are weakly perturbed~\cite{hourglass} and note that Dirac hourglass fermions are supported in both space groups $P4/ncc$ (no. 130) and $P4_2/nbc$ (no. 133)~\cite{tetragonal}, which are both minimal supergroups for the $Ccca$ structure of PtSn$_4$. A better understanding of Dirac node arcs and their bulk classification is needed, which makes finding suitable material platforms essential, and we hope that the small collection of structurally related compounds presented herein inspires research in that direction. 

Returning finally to the XMR observed in PtSn$_4$, we stress that although Dirac node arc systems, and topological semimetals in general, typically provide a platform that is favorable to XMR~\cite{niu2021materials}, they should not be assumed to have a shared origin. In the particular case of PtSn$_4$, its large carrier density and its complex multi-band Fermi surface distinguish it from other semimetals exhibiting XMR~\cite{mun2012magnetic}. Nonetheless, recent work has demonstrated the XMR of PtSn$_4$ can be understood at a semiclassical level due to electron-hole compensation~\cite{diaz2024semi} without the need to invoke band topology. As is always the case for materials showcasing XMR,  the extremely low defect concentrations and the resulting long mean free paths in PtSn$_4$ are the key ingredients that allow the XMR to be observed.

\section{Conclusions}

In this work, we have closely examined PtSn$_4$, a layered metal known for its characteristic high mobility and XMR, to disentangle these known electrical transport properties from their putative origins. We establish PtSn$_4$ as an intrinsically defect-intolerant material. Our efforts to deliberately introduce more defects through rapid cooling from a binary melt reveal that PtSn$_4$ retains excellent sample quality even when faced with adverse growth conditions.  We consider the role of the Pt and Sn sublattices as well as the dimensionality in this transport behavior, highlighting that both sublattices contribute in equal measure to the highly three-dimensional electrical properties. A comparison drawn from the resistivity measurements between the structurally similar \emph{M}Sn$_4$ (\emph{M}$=$ Rh, Ir, Pt, and Au) family shows that the defect intolerance of PtSn$_4$ does not extend to its structural analogs, which have an order of magnitude reduction in RRR among them; however, comparable sample qualities are found for isostructural and isoelectronic PdSn$_4$. The energy cost for vacancy formation, calculated via DFT, is substantially larger for PtSn$_4$ than AuSn$_4$, and is poorly compensated by relaxation, confirming that defects are not well tolerated by the PtSn$_4$ structure. STM studies further substantiate our claim with large-scale surveys revealing that Pt site defects are exceedingly rare. This natural defect intolerance when coupled to the intrinsic electron-hole compensation of the electronic band structure, is responsible for the observed XMR in PtSn$_4$. Taken together, our findings reveal a structural Goldilocks effect at play in PtSn$_4$, meaning that this particular combination of elements occupying this particular lattice are ``just right'', making the formation of defects too costly and therefore exceedingly rare.

In the world of quantum materials, ultrapure samples are highly sought after for the access they provide to intrinsic materials physics and the quality of insights they enable through measurement of properties such as quantum oscillations. At the extremely low levels of defects found in PtSn$_4$, entirely new categories of experiments become possible. When impurity scattering is sufficiently minimized, ultrapure metals can serve as platforms to study novel non-diffusive transport regimes, such as viscous and ballistic electron flow~\cite{moll2016evidence,putzke2020h,bachmann2022directional,baker2024nonlocal}. These materials also enable fundamental inquiry into the formation and functionality of disorder through the artificial creation of defects, for example by irradiation~\cite{sunko2020controlled}. Until now, much of the work in these areas has utilized PdCoO$_2$, a material whose RRRs are comparable to those of PtSn$_4$ but whose highly anisotropic transport signatures are dominated by a single (Pd) sublattice~\cite{shannon1971chemistry,tanaka1996growth,ong2010unusual,hicks2012quantum,zhang2024crystal}. In contrast, PtSn$_4$ has fully three-dimensional electrical transport and low impurity levels on both of its sublattices. Given the ease with which PtSn$_4$ cleaves, we also highlight the exciting possibility of mechanically or chemically exfoliating this material down to mono or few-layer samples, enabling the study of twisted or interfacial physics. With these advantages, PtSn$_4$ is an ideal material platform to study fundamental physics in the disorder-free limit.

\section{Methods}

\subsection{Synthesis and structural characterization}

Single crystals of slow and fast cooled PtSn$_4$ were grown out of Sn-rich self flux. The reactants Pt (99.99\,\%, ACI Metals Inc.) and Sn (99.999\,\%, ESPI Metals) in a molar ratio of 4:96 were placed into Al$_2$O$_3$ Canfield crucible sets~\cite{canfield2016use} and sealed in a quartz ampule under partial Ar atmosphere. For the slow-cooled samples (representative of the S1 phase), the setup was then slowly heated to 600\,\degree C and held at this temperature for 6 hrs followed by cooling down to 350\,\degree C over 60 hrs. At 350\,\degree C, the crystals were separated from the excess flux by centrifuging at 2000 rpm for 15 secs. For the fast-cooled samples (representative of the S2 and S3 phase), the cooling time to go from 600\,\degree C to 350\,\degree C was reduced from 60 hrs to 12 hrs and 2 hrs, respectively, followed by decanting of the excess Sn flux with a centrifuge. The as-grown crystals were layered and shiny with dimensions up to 5 $\times$ 3 $\times$ 2 mm$^3$ and were stable in the air over long-term exposure. 

A similar method of growth was followed for IrSn$_4$, AuSn$_4$ and PdSn$_4$. The growth for IrSn$_4$ was adopted and modified from a previous study \cite{tran2013observation}. For IrSn$_4$, the reactants Ir (99.95\,\%, Smart Elements) and Sn (99.999\,\%, ESPI Metals) in a molar ratio of 1:80 were placed into Al$_2$O$_3$ Canfield crucible sets~\cite{canfield2016use} and sealed in a quartz ampule under partial Ar atmosphere. The setup was then slowly heated to 900\,\degree C and held at this temperature for 12 hrs followed by cooling down to 600\,\degree C over 100 hrs, and fast finally cooling to 300\,\degree C within an hour. At 300\,\degree C, the crystals were separated from the excess flux by centrifuging at 2000 rpm for 15 secs. 

For AuSn$_4$, Au (99.999\,\%, Alpha Aesar) and Sn (99.999\,\%, ESPI Metals) were placed and sealed inside Al$_2$O$_3$ Canfield crucible sets~\cite{canfield2016use} in a molar ratio of 12:88. This was followed by a heating cycle which involved heating to 450\,\degree C over 4 hrs, holding 450\,\degree C for 12 hrs, and slow cooling down to 245\,\degree C over 72 hours before spinning with the centrifuge to separate the excess Sn flux at 245\,\degree C.

Single crystals of PdSn$_4$ were grown out of Sn-rich self flux. The reactants Pd (99.99\,\%, ACI Metals Inc.) and Sn (99.999\,\%, ESPI Metals) in a molar ratio of 2:98 were placed into Al$_2$O$_3$ Canfield crucible sets~\cite{canfield2016use} and sealed in a quartz ampule under partial Ar atmosphere. The setup was then slowly heated to 600\,\degree C and held at this temperature for 6 hrs followed by cooling down to 250\,\degree C over 85 hrs. At 250\,\degree C, the crystals were separated from the excess flux by centrifuging at 2000 rpm for 15 secs.

The crystal structure and orientation of the crystals were verified using X-ray diffraction (XRD) using a Bruker D8 Advance with Cu K$\alpha$1 radiation ($\lambda$ = 1.54056 \AA). For powder XRD, the mechanically ground sample was annealed in vacuum at 450\,\degree C for 12 hrs to restore the crystallinity. The elemental composition and homogeneity of the samples were determined through energy dispersive X-ray spectroscopy (EDX) using a Philips XL30 scanning electron microscope equipped with a Bruker Quantax 200 energy dispersion x-ray microanalysis system and an XFlash 6010 SDD detector, and a Tescan Amber scanning electron microscope with Oxford AZtec software and Ultime Max 170 detector.

\subsection{Electrical transport measurements}
Electrical transport or resistivity measurements were performed using the Quantum Design Physical Property Measurement System (PPMS) with two kinds of contact geometries for the in-plane and out-of-plane directions, respectively. A conventional four-probe geometry was used for the in-plane measurements, where Pt (50 $\mu$m) leads were directly attached to a sample with silver epoxy. For the out-of-plane measurements, a modified U-shaped four-probe geometry was employed \cite{bud1998anisotropic}. This geometry involved the current leads being placed on the top and bottom layer of a cut square-shaped sample in a U shape with dot voltage leads placed in the middle of the layers. The targeted U-shaped contacts were achieved by masked evaporation of Au electrodes onto the top and bottom layers of the sample, followed by attaching the Pt wires to the Au-deposited sides with silver epoxy for current and voltage channel connection to the PPMS transport puck.

\subsection{Scanning tunneling microscopy}
Single crystals of PtSn$_4$ were cleaved in-situ at room temperature and in an ultrahigh-vacuum preparation chamber. After cleaving, the samples were directly transferred into the scanning tunneling microscope (STM) head. Chemically etched and annealed tungsten tips were used, the tip is characterized by scanning on the coin gold sample next to the PtSn$_4$ samples. The STM measurements were carried out using a commercial CreaTec STM at 4.2~K. 

\subsection{DFT calculations} 
Electronic structure calculations were performed within the framework of density functional theory (DFT) as implemented in the VASP package~\cite{vasp1, vasp2, vasp3}. 
The generalized gradient approximation with the PBE parameterization for the exchange-correlation functional~\cite{perdew1996generalized, PBE} was used. The relaxations have been computed in $2\times 2\times 2$ supercells for PdSn$_4$, PtSn$_4$ and AuSn$_4$ and in $2\times 2\times 1$ supercells for IrSn$_4$ (due to its doubled out-of-plane lattice parameter) at the $\Gamma$-point only. The pymatgen package was used to construct the first Brillouin zone and path parameterizations therein~\cite{pymatgen}.

\section*{Data availability}
The data that support the findings of this study are available from the corresponding author upon reasonable request.

\section*{Acknowledgements}

We gratefully acknowledge technical support from Miles Brodie, James Day, Pinder Dosanjh, Jacob Kabel, Anita Lam, Xiyang Li, Mohamed Oudah, and Bobby Lin. This work was supported by the Natural Sciences and Engineering Research Council (NSERC) of Canada, the CIFAR Quantum Materials program, and the Sloan Research Fellowships program. This research was undertaken thanks in part to funding from the Max Planck-UBC-UTokyo Centre for Quantum Materials and the Canada First Research Excellence Fund, Quantum Materials and Future Technologies Program. NH acknowledges financial support from the Max Planck Institute for Solid State Research in Stuttgart, Germany, and the Deutsche Forschungsgemeinschaft (DFG, German Research Foundation) - TRR 360 - 492547816.

\section*{Author contributions}

Crystal growth was performed by S.S. and A.K.S. Powder x-ray diffraction and electrical transport measurements were performed by S.S. Scanning tunneling microscopy measurements were performed by D.C., A.N.W., and M.A. under the guidance of D.A.B and S.A.B. Density functional theory calculations were performed by N.H. Single crystal x-ray diffraction was performed by A.K.S. The project was supervised by A.M.H. and S.A.B. The manuscript was written by S.S., D.C., N.H., and A.M.H. with input from all authors.\\

\section*{Competing interests}
The authors declare no competing interests.

\bibliography{refs}

\end{document}